\documentclass[showpacs,aps,amssymb]{revtex4}
\usepackage{bm}
\usepackage{graphicx}

\begin{document}

\def\ba{{\bf a}}
\def\bk{{\bf k}}
\def\bp{{\bf p}}
\def\bq{{\bf q}}
\def\br{{\bf r}}
\def\bv{{\bf v}}
\def\bx{{\bf x}}
\def\P{{\bf P}}
\def\bR{{\bf R}}
\def\bK{{\bf K}}
\def\la{\langle}
\def\ra{\rangle}
\def\beq{\begin{equation}}
\def\eeq{\end{equation}}
\def\bea{\begin{eqnarray}}
\def\eea{\end{eqnarray}}
\def\bdm{\begin{displaymath}}
\def\edm{\end{displaymath}}

\title{Transition to the Giant Vortex State in an Harmonic Plus
Quartic Trap}
\author{H. Fu and E. Zaremba}
\affiliation{Department of Physics, Engineering Physics and
Astronomy, Queen's University, Kingston, Ontario
 K7L 3N6, Canada.}
\date{\today}
\begin{abstract}
We consider a rapidly rotating Bose-condensed gas in an harmonic
plus quartic trap. At sufficiently high rotation rates the
condensate acquires an annular geometry with the superposition
of a vortex lattice. With increasing rotation rate the lattice
evolves into a single ring of vortices. Of interest is the
transition from this state to the giant vortex state in which
the circulation is carried by only a central vortex. By
analyzing the Gross-Pitaevskii energy functional variationally,
we have been able to map out the phase boundary between these two
states as a function of the rotation rate and the various
trapped gas parameters. The variational results are in good
qualitative agreement with those obtained by means of a direct
numerical solution of the Gross-Pitaevskii equation.

\end{abstract}
\maketitle

\section{Introduction}

In a recent paper, Fetter {\it et al.}~\cite{fetter05} investigated 
the vortex structure
in a rapidly rotating Bose condensate confined by an harmonic plus
quartic trapping potential. The interest in quartic confinement stems
from the fact that the rotation rate $\Omega$ is limited in harmonic
traps by the trap frequency $\omega_\perp$, at which point the
centrifugal potential destabilizes the system and the radius
and angular momentum of the condensate diverge. The addition of the 
quartic potential~\cite{fetter01}
avoids this instability and allows one to investigate
rotation rates higher than $\omega_\perp$. In this regime,
the condensate exhibits a more
complex structure, both with regard to its density distribution and the
arrangement of vortices within it~\cite{lundh02,kasamatsu02,fetter05}. 

The combined harmonic, centrifugal and
quartic potentials give rise to a Mexican hat potential and the mean
density acquires a local minimum on the axis of symmetry of the trap. 
Within a Thomas-Fermi analysis, the central density goes to zero at some
limiting angular velocity
$\Omega_h$~\cite{fetter05,kavoulakis03,fischer03}, 
beyond which the condensate takes on an annular structure. This 
behaviour is also found in numerical solutions of the Gross-Pitaevskii 
equation~\cite{fetter05,lundh02,kasamatsu02}.
Within some range of
interaction strengths and angular velocities, these calculations reveal
a structure in which a ring of vortices surrounds a central hole
containing a multiply quantized vortex. With increasing rotation rate,
a new state is favoured in which the annular condensate is vortex free
and all the circulation is carried by a central vortex. This state
with a multiply quantized central vortex is referred to as the giant
vortex state~\cite{kasamatsu02,fischer03,kavoulakis03}.   

In this paper we are concerned with the transition from the state
with a single ring of vortices, which we also refer to as an 
annular array,
to the giant vortex state. For low interaction strengths, the
numerical~\cite{fetter05,lundh02,kasamatsu02} and
analytical~\cite{jackson04}
evidence indicates that the transition is continuous, with the radius of
the ring shrinking in size until the ring is absorbed by the central 
hole.
The situation at higher
interaction strengths seems to be different, with the annular array 
persisting as a metastable state. In any case, one
expects the giant vortex state to be preferred energetically
above some critical angular velocity $\Omega_c$. We determine the phase
boundary between these two states by variationally minimizing the
Gross-Pitaevskii (GP) energy of the annular array and comparing this
with the corresponding energy of the giant vortex state. Our results are
in essential agreement with those of Kim and Fetter~\cite{kim05} which
are based on the same physical model but are obtained using a different
calculational approach.

In Sec. II we present the physical model of the annular array within the
context of the Gross-Pitaevskii energy functional. The nature of the
problem suggests that a good starting point for the calculation of the 
GP energy can be achieved by replacing the annular array by a vortex
sheet. Nevertheless, the discrete nature of the vortex cores leads to
important corrections to the energy which determine the relative
stability of the annular array and giant vortex states.
We thus begin with a detailed analysis of the vortex 
sheet problem in Sec. III and then examine in Sec. IV the various 
corrections to the vortex sheet energy due to the vortex cores. 
Our numerical results are 
presented in Sec. V and we conclude with a discussion in Sec. VI.
\section{Energy}
The GP energy functional in a frame of reference rotating with angular
velocity $\Omega$ is given by~\cite{fetter05}
\begin{equation}
E[\Psi] = \int d^3r \left \{ {\hbar^2 \over 2M} |\nabla \Psi|^2
+ V|\Psi|^2 + {2\pi a \hbar^2 \over M} |\Psi|^4 \right \} -
\Omega L_z\,,
\label{GPEF}
\end{equation}
where $\Psi$ is the condensate wave function, $a>0$ is the
$s$-wave scattering length and $L_z = \int d^3r \Psi^* \hat {\bf
z} \cdot (\br \times {\bf p}) \Psi$ is the $z$-component of the
angular momentum. The normalization of the wave function is $N =
\int d^3r |\Psi|^2$, where $N$ is the total number of particles.
As in previous
discussions~\cite{fetter05,kasamatsu02,kavoulakis03,kim05}, we consider
the two-dimensional limit in which the
confining potential is taken to be an
harmonic plus quartic potential in the radial direction with no
confinement in the $z$-direction:
\begin{equation}
V(r) = {1\over 2} M\omega_\perp^2\left ( r^2 + \lambda {r^4
\over d_\perp^2} \right ) = {1\over 2} \hbar \omega_\perp \left
({r^2 \over d_\perp^2} + \lambda  {r^4 \over
d_\perp^4}\right )\,.
\label{potential}
\end{equation}
Here $d_\perp = \sqrt{\hbar/M\omega_\perp}$ is the harmonic
oscillator length. For this two-dimensional situation the wave function 
$\Psi$ depends only on the radial and angular variables $r$ and 
$\phi$, respectively.

\begin{figure}[h]
\scalebox{.6}
{\includegraphics[0,0][312,312]{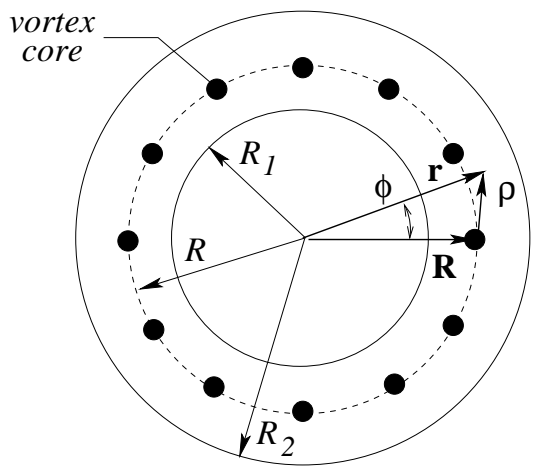}}
\caption{Schematic of the annular array configuration.}
\label{ring}
\end{figure}

Our purpose is to determine the equilibrium state of the system by
minimizing the GP energy functional for a given angular velocity
$\Omega$. Variations with respect to $\Psi$ yield the GP equation which
can be solved numerically to determine the exact equilibrium state.
Solutions of this kind have been
obtained~\cite{fetter05,lundh02,kasamatsu02} but the calculations are
numerically demanding and cannot be performed in all parameter regimes.
It is then useful to adopt a variational approach which has the
added advantage of providing more physical insight into the
nature of the solutions. To this end we
introduce the amplitude-phase representation of the wave
function, $\Psi = \sqrt{\sigma n} e^{i\theta}$, where $\sigma$
is the density per unit length in the $z$-direction. The 
two-dimensional density $n(r,\phi)$ then has the normalization
\begin{equation}
\int d^2r\, n(\br) = 1\,.
\label{norm}
\end{equation}
Using $d_\perp$ as the unit of length and $\hbar \omega_\perp$
as the unit of energy, the (dimensionless)
energy per particle takes the form
\begin{equation}
E[n,\bv] = \int d^2r \left \{ {1 \over 2} |\nabla
\sqrt{n}|^2 + {1\over 2} nv^2 + Vn + {1\over 2} gn^2 - \Omega
\cdot \br \times \bv n \right \}
\label{E_nv}
\end{equation}
where the condensate velocity is given by $\bv = \nabla \theta$
and the interaction strength is $g = 4\pi \sigma a$. In this context 
the equilibrium state is determined by the pair of functions $n$ and
$\bv$ which minimize this energy functional. Such an approach is
feasible when, as in the present case, the physical states of interest
are known. 
For $\Omega > \Omega_h$ the condensate forms an annulus with
some distribution of vortices~\cite{fetter05,kavoulakis03}. Initially
the vortices are arranged on a regular lattice, but as the central hole 
becomes well-established, the lattice evolves into concentric 
rings of vortices around the central hole. Eventually, a single ring 
is formed 
and the transition from this state to the giant vortex state is the 
transition of
particular interest here. The geometry and variables used to
describe the annular array are illustrated in Fig.~\ref{ring}.

The velocity field for $N_r$ vortices on a 
ring of radius $R$ is assumed to be given by a superposition of
the velocity fields of individual vortices,
\begin{equation}
\bv(\br) = \sum_{j=1}^{N_r} {\hat {\bf z} \times (\br - \bR_j)
\over |\br - \bR_j|^2} + {N_0 \over r}\hat {\bm{\phi}\,,}
\label{velocity}
\end{equation}
where $\bR_j$ are the positions of the singly-quantized vortices. 
The second term accounts for the vorticity of a central vortex
of strength $N_0$. In this approach, the form of the velocity field is
effectively fixed,
but $R$, $N_r$ and $N_0$ remain as variational parameters.
For the ring configuration of Fig.~\ref{ring}, 
the velocity and density have an angular
periodicity of $2\pi/N_r$, for example,
\begin{equation}
\bv(r,\phi+2\pi/N_r) = \bv(r,\phi)\,,
\end{equation}
and all such quantities can be expanded in a Fourier series. We write
\begin{equation}
\bv(\br) = \bv_0(r) + \bv_1(\br)
\label{v0v1}
\end{equation}
where $\bv_0$ is the azimuthally averaged part of the velocity
field which is given by~\cite{kim04}
\begin{equation}
\bv_0(r) = \left [ {N_0 \over r} \theta(R-r) + {N_0 +N_r \over r}
\theta(r-R) \right ] \hat {\bm{\phi}}\,.
\label{v0}
\end{equation}
This can be interpreted as the velocity field of a central vortex and a 
vortex sheet of radius $R$ across which the velocity is
discontinuous. It should be emphasized that this part of the velocity
field is perfectly general and is independent of the assumed vortex
superposition in Eq.~(\ref{velocity}). It is valid even for the exact GP
solution having a net circulation of $N_0$ in the low-density hole
region and $N_r$ vortex singularities arranged on a ring of radius $R$.

The remaining part $\bv_1 = v_{1r} \hat \br +
v_{1\phi} \hat {\bm{\phi}}$ has the Fourier expansion~\cite{kim04}
\begin{equation}
v_{1r}(\br) = -{N_r \over r} \sum_{k=1}^{N_r} \sin kN_r\phi
\left ({r_<\over r_>}\right )^{kN_r}\,,
\label{v1r}
\end{equation}
\begin{equation}
v_{1\phi}(\br) = {\rm sgn}(r-R) {N_r \over r} \sum_{k=1}^{N_r} 
\cos kN_r\phi \left ({r_<\over r_> }\right )^{kN_r}\,,
\label{v1phi}
\end{equation}
where $r_<$ ($r_>$) is the lesser (greater) of $r$ and $R$.
It is this part which contains the singular behaviour of the velocity
field at the positions of the vortices. For example, by performing the
sum in Eq.~(\ref{v1phi}) we have
\begin{equation}
v_{1\phi} = {\rm sgn}(r-R){N_r \over r} {\beta (\cos N_r \phi - \beta) 
\over 1+\beta^2 -2\beta \cos N_r \phi}\,,
\label{v1phi_2}
\end{equation}
where $\beta = (r_</r_>)^{N_r}$.
One can easily see from this expression that $v_{1\phi} \propto
\rho^{-1}$ near each vortex, where $\rho$ is the distance from
the vortex. We also note that the $\bv_1$ velocity field is strongly
localized near $r =R$ since the factor $\beta$ 
decreases rapidly with increasing $|r-R|$ when $N_r$ is large. 
That is, the velocity field rapidly approaches the long-range part
described by $\bv_0$ as one moves away from the ring of vortices, and is
therefore essentially azimuthal at the edges of the annulus. In a sense,
this provides an {\it a posteriori} justification for the validity of 
the superposition of individual vortex velocity fields in 
Eq.~(\ref{velocity})
since no geometric constraint is being imposed on the flow by 
the boundaries of the annulus. This is unlike the
situation of a single vortex near the surface of a planar boundary where
the velocity field is strongly modified from that of an isolated vortex
by the presence of the boundary~\cite{anglin02}.

Using Eq.~(\ref{v0v1}), the classical kinetic energy
consists of the following terms:
\begin{equation}
E_K = {1\over 2} \int d^2r \, n_0(r) v_0^2(r) + \int d^2r
\,n(\br) v_0(r) v_{1\phi}(\br) + {1\over 2} \int d^2r \,n(\br)
v_1^2(\br)\,,
\label{E_K}
\end{equation}
where
\begin{equation}
n_0(r) \equiv {1\over 2\pi} \int_0^{2\pi} n(r,\phi) \,d\phi\,.
\end{equation}
We see that the separation of the velocity field into $\bv_0$
and $\bv_1$ leads to a kinetic energy contribution which only
involves $\bv_0$ and the angular average of the density $n_0(r)$.
It is thus natural to decompose the density as 
$n(\br) = n_0(r) + n_1(\br)$, where $n_1$ is the part
containing all higher Fourier components. 

With these definitions, the energy functional can be written as
\begin{equation}
E[n,\bv] = E_{VS}[ n_0, \bv_0] + E_{VC}[n,\bv]\,,
\end{equation}
with 
\begin{equation}
E_{VS}[n_0, \bv_0] = \int d^2r\, \left [ {1\over 2} n_0
v_0^2 + Vn_0 + {1\over 2} g n_0^2 - \Omega r n_0 v_0
\right ] 
\label{E_VS}
\end{equation}
and
\begin{equation}
E_{VC}[n,\bv] = \int d^2r \, \left [ {1 \over 2} |\nabla
\sqrt n|^2 + {1\over 2} n v_1^2 + nv_0 v_{1\phi}  
- \Omega rn v_{1\phi} + {1\over 2} g n_1^2 \right ]\,.
\label{E_VC}
\end{equation}
The term $E_{VS}$ signifies the contribution to the energy that
arises when the discrete array of vortices is replaced by a vortex 
sheet (VS) of radius $R$ and circulation $N_r$. We note that
the variables entering this part of the energy functional are
independent of the angular variable $\phi$. We identify the second term
$E_{VC}$ as the vortex core (VC) energy. If this contribution to the
energy is 
neglected, the $E_{VS}$ functional by itself is sufficient to define
both $n_0$ and $v_0$. At this level of approximation the functional
provides a Thomas-Fermi (TF) description of the vortex sheet 
configuration.
We emphasize, however, that no approximations have been made in
writing the energy functional as the sum of the two separate
contributions in Eqs.~(\ref{E_VS}) and (\ref{E_VC}).

The second contribution $E_{VC}$ accounts for the actual
vortex core structure and depends on the full inhomogeneous
nature of the density and velocity field near each vortex. 
We note that the singular part of the velocity field
arises in the terms containing $\bv_1$. This singularity
is of course compensated by the density which must behave as $n
\propto \rho^2$ near the core of each vortex. To represent this
behaviour we write the density as~\cite{footnote1}
\begin{equation}
n(\br) = F(\br) \tilde n(\br)\,,
\label{density}
\end{equation}
where $F(\br)$ is an envelope function which accounts for the
depletion of the density near each vortex with respect to some
overall smooth background density $\tilde n(\br)$. Since $n(\br)$ is
normalized to unity, $\tilde n(\br)$ in general will not be.

The factorization of the density in Eq.~(\ref{density}) is not unique
but it allows for a convenient variational representation of the 
density depletion around each vortex that occurs
on a length scale $\xi$ which is small in comparison to all other
characteristic lengths in the problem. To be specific we choose
an envelope function having the form
\begin{equation}
F(\br) = \sum_{i=1}^{N_r} e^{-|\bR_i - \bR_1|^2/\xi^2} -
\sum_{i=1}^{N_r} e^{-|\br -\bR_i |^2/\xi^2}\,,
\label{envelope}
\end{equation}
which gives the density depletion a gaussian profile.
The parameter $\xi$ represents the vortex core radius and is
treated as a variational parameter. It will be of the order of
the local healing length  and we anticipate that $\xi \ll b$,
where $b = 2\pi R/N_r$ is the inter-vortex spacing. In this
situation, the density close to each vortex is given to a good
approximation by $n(\br) \simeq (1-\exp(-\rho^2/\xi^2)) \tilde
n(\br)$, which has the required $n \propto \rho^2$ behaviour. An
alternative to Eq.~(\ref{envelope}) is to represent each vortex core by 
the piece-wise continuous function
$f(\rho) = (\rho/\xi)^2$ for $\rho < \xi$ 
and $f(\rho) = 1$ for
$\rho > \xi$~\cite{fischer03,kim05}. The advantage of
the gaussian core is that it allows an essentially analytic calculation
of the core energy. However, in view of the variational nature of the 
calculation, one would expect the two forms 
to yield quantitatively similar results.

Once the core structure is accounted for via the envelope
function $F(\br)$, the background density $\tilde n(\br)$ is
expected to have a weak angular variation. It can itself be treated 
as a variational function in the minimization of the total energy.
However, we expect and confirm that the VC contribution to the 
energy is relatively small in comparison to $E_{VS}$ and as such 
can be treated as a perturbation. Our strategy is therefore to 
minimize $E_{VS}$ with respect to $n_0$ and $v_0$ and to use the 
information provided by this minimization in the evaluation of 
$E_{VC}$.  Since the $\bv_1$ velocity field is highly
localized near $r=R$, the terms containing $\bv_1$ in Eq.~(\ref{E_VC})
are only sensitive to the background density in this region.
Given its smooth angular variation, we will for simplicity 
choose $\tilde n(r)$ to be a function of 
only the radial variable $r$. According to its definition in
Eq.~(\ref{density}),
this implies that the background density is related to the vortex sheet
density by
\begin{equation}
\tilde n(r) = {n_0(r) \over F_0(r)}\,,
\label{n_tilde}
\end{equation}
where $F_0(r)$ is the angular average of $F(\br)$. We use this
equation to estimate $\tilde n$ near $r=R$.
In this way all contributions to $E_{VC}$ become functions 
of $\xi$ and to complete the calculation, the VC
energy is minimized with respect to this parameter.
\section{The Vortex Sheet Approximation}
\label{sheet}

In this section we calculate the energy within the vortex sheet
approximation (VSA) obtained by minimizing Eq.~(\ref{E_VS}) with 
respect to the density $n_0$ and the circulation parameters of 
the azimuthally averaged velocity given by Eq.~(\ref{v0}).
Here we denote the circulation parameters by $\nu_1 = N_0$ and
$\nu_2 = N_0 +N_r$. It is reasonable to treat these parameters 
as continuous if the circulation is large. 
To impose the normalization constraint on the density we
minimize the free energy $F_{VS} = E_{VS} -\mu\int d^2r\, n_0$ where 
$\mu$ is the chemical potential. Variation of $F_{VS}$ with respect 
to $n_0$ gives
\begin{equation}
gn_0(r) = \left \{ \begin{array}{c} \mu_1 - U_1(x),\quad r<R\,\,
\\ \mu_2 - U_2(x),\quad r>R\,, \end{array} \right .
\label{den_TF}
\end{equation}
where
\begin{equation}
\mu_i = \mu + \Omega \nu_i\,,
\label{mu_i}
\end{equation}
and
\begin{equation}
U_i(x) = {1 \over 2} \left ( {\nu_i^2 \over x} + x + \lambda x^2
\right ) 
\label{U_i}
\end{equation}
with $x = r^2$. The inner and outer radii, $R_i=\sqrt{x_i}$, of the 
annulus are defined by 
\begin{equation}
\mu_i = U_i(x_i),\quad i=1,2\,.
\label{x_i}
\end{equation}
By its definition, $n_0$ is the azimuthally averaged density and
is therefore a continuous function of the radial variable $r$. 
The requirement that $n_0$ be continuous across the vortex sheet
leads to the relation
\begin{equation}
R\Omega  = {1\over 2} \left ( {\nu_1\over R} +{\nu_2\over R}
\right )\,,
\label{bc}
\end{equation}
which indicates that the average of the velocity on either
side of the vortex sheet is equal to the velocity for rigid
body rotation at the radius $r=R$. 
Eq.~(\ref{bc}) can also be interpreted in terms of
the circulation provided by a uniform vortex lattice. The number of
vortices contained within a circle of radius $R$ that is needed to
give the velocity $R\Omega$ at $r=R$ is $\nu_L = R^2\Omega$. Thus, 
Eq.~(\ref{bc}) is equivalent to the relation $\nu_L = 
(\nu_1+\nu_2)/2$.

The density given by Eq.~(\ref{den_TF}) also depends on the
chemical potential $\mu$. For a given $\nu_1$ and $\nu_2$, this
parameter is fixed by the normalization of the density in
Eq.~(\ref{norm}). This gives the additional relation
\begin{equation}
g = \pi \int_{x_1}^{x_0} dx\,(\mu_1 - U_1(x)) + \pi
\int_{x_0}^{x_2} dx\,(\mu_2 - U_2(x))\,, 
\label{g_norm}
\end{equation}
where $x_0 = R^2$.

Using Eq.~(\ref{den_TF}), the free energy can be expressed as
\begin{equation}
F_{VS} = -{\pi \over 2g} \int_{x_1}^{x_0} dx\,(\mu_1 - U_1(x))^2
-{\pi \over 2g} \int_{x_0}^{x_2} dx\,(\mu_2 - U_2(x))^2\,.
\label{F_VS}
\end{equation}
The equilibrium state corresponds to the minimization of $F_{VS}$
with respect to the two parameters $\nu_1$ and $\nu_2$. Noting
that $x_0$, $x_1$ and $x_2$ are implicitly functions of $\nu_1$
and $\nu_2$, this variation yields the equations
\begin{eqnarray}
&&\Omega \int_{x_1}^{x_0} dx\,(\mu_1 - U_1(x)) =
\nu_1 \int_{x_1}^{x_0} {dx\over x}\,(\mu_1 - U_1(x))\,,\label{F_min_a}\\
&&\Omega \int_{x_0}^{x_2} dx\,(\mu_2 - U_2(x)) =
\nu_2 \int_{x_0}^{x_2} {dx\over x} \,(\mu_2 - U_2(x)) \,.
\label{F_min_b}
\end{eqnarray}
The six nonlinear equations, Eqs.~(\ref{x_i}-\ref{g_norm}),
(\ref{F_min_a}) and
(\ref{F_min_b}), are sufficient to determine 
the six unknown parameters $\nu_1$, $\nu_2$, $R$, $R_1$, $R_2$
and $\mu$.

The procedure described corresponds to a free variation of the
circulation parameters $\nu_1$ and $\nu_2$. Alternatively,
the minimization can be carried out with imposed constraints,
such as fixing
the radius $R$ of the sheet or the number of vortices $N_r$ in the 
ring. In the first case, the pair of equations (\ref{F_min_a}) and
(\ref{F_min_b}) is
replaced by the single equation ($R$ fixed)
\begin{eqnarray}
&&\Omega \int_{x_1}^{x_0} dx\,(\mu_1 - U_1(x)) - \Omega
\int_{x_0}^{x_2} dx\,(\mu_2 - U_2(x)) + 2x_0 \Omega 
\int_{x_0}^{x_2} {dx\over x} \,(\mu_2 - U_2(x)) \cr
&&\hskip 1.5truein
=\nu_1 \left \{\int_{x_1}^{x_0} {dx\over x} \,(\mu_1 - U_1(x)) +
\int_{x_0}^{x_2} {dx\over x} \,(\mu_2 - U_2(x)) \right \}\,,
\label{R-fixed}
\end{eqnarray}
while in the second ($N_r$ fixed) we have
\begin{eqnarray}
&&\Omega \int_{x_1}^{x_0} dx\,(\mu_1 - U_1(x)) + \Omega
\int_{x_0}^{x_2} dx\,(\mu_2 - U_2(x)) - N_r 
\int_{x_0}^{x_2} {dx\over x} \,(\mu_2 - U_2(x)) \cr
&&\hskip 1.5truein
=\nu_1 \left \{\int_{x_1}^{x_0} {dx\over x} \,(\mu_1 - U_1(x)) +
\int_{x_0}^{x_2} {dx\over x} \,(\mu_2 - U_2(x)) \right \}\,.
\label{N_r-fixed}
\end{eqnarray}
In the limit $N_r \to 0$, ($\nu_1 = \nu_2$) the latter equation
reduces to the minimization equation for the giant vortex state:
\begin{equation}
\Omega \int_{x_1}^{x_2} dx\,(\mu_1 - U_1(x)) 
=\nu_1 \int_{x_1}^{x_2} {dx\over x} \,(\mu_1 - U_1(x))\,.
\end{equation}
The parameter $x_0$ of course has no significance in this limit.
We later make use of some of these constrained minimizations.

Although the above equations can be solved numerically (to be
described later), it is useful to perform a perturbative
analysis in order to gain some insight into the form of the
solutions. For $\Omega > \Omega_h$, the
condensate is confined to an annulus whose radius increases with
$\Omega$. At the same time, the width of the annulus decreases.
We therefore expect the parameter $w = x_2 -x_1$ to be small in
comparison to $x_0$. In this situation, we can find useful
relations between the various parameters by expanding the
integrals in Eq.~(\ref{F_min_a}) in a Taylor series,
\begin{equation}
\int_{x_1}^{x_0} dx\, f(x) = f(x_1) \Delta_1 + {1\over 2}
f'(x_1)\Delta_1^2 + {1\over 6} f''(x_1) \Delta_1^3 + \cdots\,,
\end{equation}
where $\Delta_1 \equiv x_0 - x_1$. In our case, $f(x_1) = 0$ and
the expansion starts with terms of order $\Delta_1^2$. The
expansion of the integrals in 
Eq.~(\ref{F_min_a}) can thus be viewed as providing a power
series expansion for $\nu_1$:
\begin{equation}
\nu_1 = \nu_1^{(0)} + \nu_1^{(1)}\Delta_1 +\cdots \,.
\end{equation}
We insert this expansion for $\nu_1$ wherever it appears in
Eq.~(\ref{F_min_a}), including the $\mu_1$ and $U_1(x)$ terms,
and generate a power series in  $\Delta_1$ for each side of the
equation. Equating the coefficients of like powers of $\Delta_1$
and retaining terms to order $\Delta_1^4$,  we find
\begin{equation}
\nu_1 = \Omega x_1\left [ 1 + {2\over 3} {\Delta_1\over x_1} -
{\alpha_1 \over 18} \left ( {\Delta_1\over x_1} \right )^2
+\cdots\right ] \,,
\label{nu_1}
\end{equation}
where
\begin{equation}
\alpha_1 = {1-2\Omega^2 + \lambda x_1 \over 1-\Omega^2 +
2\lambda x_1}\,.
\end{equation}
Analyzing Eq.~(\ref{F_min_b}) in a similar way, we obtain
\begin{equation}
\nu_2 = \Omega x_2\left [ 1 - {2\over 3} {\Delta_2\over x_2} -
{\alpha_2 \over 18} \left ( {\Delta_2\over x_2} \right )^2
+\cdots\right ] \,,
\label{nu_2}
\end{equation}
where $\Delta_2 \equiv x_2-x_0$ and $\alpha_2$ is obtained from
the expression for $\alpha_1$ by replacing $x_1$ by $x_2$.
These relations show that
\begin{equation}
x_0 = {1\over 2\Omega}(\nu_1+\nu_2) = {1\over 2}(x_1+x_2) -
{1\over 12} \left ( {\alpha_1\over x_1} \Delta_1^2 +
{\alpha_2\over x_2} \Delta_2^2 \right ) +\cdots \,,
\end{equation}
that is, $R^2 =(R_1^2+R_2^2)/2$ with corrections of order
$\Delta_i^2$. These equations also imply that $\Delta_i \simeq w/2$ 
to lowest order.

To obtain an explicit expression for $x_1+x_2$ we take the
difference of the two equations in Eq.~(\ref{x_i}), thereby eliminating
$\mu$. Using the above expansions in powers of $w$, we then find
$x_1+x_2 \simeq (\Omega^2-1)/\lambda$, which implies
\begin{equation}
x_0 = {\Omega^2 - 1 \over 2\lambda},
\label{x_0}
\end{equation}
with corrections of order $(w/x_0)^2$. This leading order result
for $x_0$ is the same as the expression for $(x_1+x_2)/2$
obtained for a vortex lattice with a hole~\cite{fetter05}. The
reason for this can be seen by inserting the lowest order result
$\nu_i = \Omega x_i$ into Eq.~(\ref{x_i}). This gives
$\mu={1\over 2} \left ( x_i(1-\Omega^2)+\lambda x_i^2 \right )$
which is recognized as the equation giving the boundaries of the
annulus in the vortex lattice state. This correspondence
indicates that the annular array in
the large-$\Omega$ limit is effectively undergoing rigid body
rotation.

The result for $x_0$ in Eq.~(\ref{x_0}) shows that the denominator
of $\alpha_1$ is $1-\Omega^2+2\lambda x_1 \simeq -\lambda w$.
Thus the terms formally of order $\Delta_i^2$ in Eqs.~(\ref{nu_1}) and
(\ref{nu_2}) in fact give rise to corrections to $\nu_i$ that
are of order $w$. From Eqs.~(\ref{nu_1}) and (\ref{nu_2}) we
thus find
\begin{equation}
N_r = \nu_2-\nu_1 \simeq {1\over 3}\Omega w \left [ 1+{1\over 4}
{\Omega^2 - 1/3 \over \Omega^2 -1} + \cdots \right ] \,.
\label{N_r}
\end{equation}

Another useful relation follows from the 
expansion of the normalization condition, Eq.~(\ref{g_norm}). To order
$w^3$ we have
\begin{equation}
g = {\pi \over 8} \left [ \left . {\partial U_2 \over 
\partial x} \right |_{x_2}  - \left . {\partial U_1 \over 
\partial x} \right |_{x_1} \right ]w^2 - {\pi \over 48}
\left [ \left . {\partial^2 U_2 \over
\partial x^2} \right |_{x_2}  + \left .
{\partial^2 U_1 \over \partial x^2} \right |_{x_1}
\right ]w^3\,.
\end{equation}
The first term on the right hand side appears to be of order $w^2$, but
if the potential derivatives in
this term are evaluated to lowest order in $w$ we obtain
$\lambda(x_2-x_1)$, which in fact is of order $w$. Thus, to include all
contributions of order $w^3$, the potential
derivatives in the first term must be expanded to first order in
$w$ and the second term must also be retained. Doing so one finds
\begin{equation}
w = \left ( {12g \over \pi(\lambda + \Omega^2/2x_0)} \right )^{1/3}
= \left ( {12g \over \pi\lambda(1 + \Omega^2/(\Omega^2-1))} 
\right )^{1/3}\,.
\end{equation}
For large $\Omega$ we have $w\simeq (6g/\pi\lambda)^{1/3}\equiv
w_\infty$. Since $w = R_2^2 - R_1^2$, the physical width of the annulus
is given by
\begin{equation}
d\equiv R_2-R_1 = \sqrt{{\lambda \over 2}}{w_\infty \over \Omega} \left
(1+ {1\over 3\Omega^2}+\cdots \right )\,,
\end{equation}
that is, $d \propto \Omega^{-1}$ for large $\Omega$
whereas the radius $R$ of the vortex sheet is proportional to
$\Omega$.
We also note that the strength of the central vortex is $N_0 =
\Omega x_0 - N_r/2
\simeq \Omega^3/2\lambda$ for large $\Omega$, while that of 
the vortex sheet is $N_r \simeq 5\Omega w_\infty/12$.

To obtain accurate results for all $\Omega$ we solve
Eqs.~(\ref{x_i}-\ref{g_norm}), (\ref{F_min_a}) and (\ref{F_min_b})
numerically. We have found that this can be done in a
straightforward iterative manner. We start by using the
large-$\Omega$ expressions $\nu_1 = \Omega^3/2\lambda$, $\nu_2=
\nu_1+5\Omega w_\infty/12$, $x_0 = (\Omega^2-1)/2\lambda$ and $x_2
= x_0+w_\infty/2$. We then estimate an approximate 
chemical potential as $\mu' =
U_2(x_2) - \Omega\nu_2$, define $\mu_1 = \mu'+\Omega\nu_1$ and
determine $x_1$ from $U_1(x_1) = \mu_1$. We now calculate
the integrals $F_i(a,b) = \int_a^b dx\,(\mu_i-U_i(x))$ and
$G_i(a,b) = \int_a^b dx\,(\mu_i-U_i(x))/x$ in order to
determine new values of the circulations from Eqs.~(\ref{F_min_a}) and
(\ref{F_min_b}): 
$\nu_1 = \Omega F_1(x_1,x_0)/G_1(x_1,x_0)$ and $\nu_2 =\Omega
F_2(x_0,x_2)/G_2(x_0,x_2)$. At the same time we use Eq.~(\ref{g_norm})
to determine the
effective coupling strength $g' = \pi(F_1(x_1,x_0)+F_2(x_0,x_2))$.
The chemical potential which gives the desired value of $g$ is
$\mu = \mu'+\Delta \mu$ and the change in chemical potential is
estimated as $\Delta \mu = (g-g')/\pi(x_2-x_1)$. With these
updated values of $\nu_1$, $\nu_2$ and $\mu$ we recalculate
$x_0=(\nu_1+\nu_2)/2\Omega$ and $x_i$ from $U(x_i) = \mu_i$, and 
repeat the remaining steps described. Convergence to an accuracy of 
one part in $10^6$ typically required 10-20 iterations.

\begin{table}[here]
\begin{center}
\begin{tabular*}{140mm}{@{\extracolsep{\fill}}ccccccccc} \hline
$\Omega$&$N_0$&$N_r$&$R_1$
&$R_2$\,&$R$&$d$&$b$&$\xi_{\rm min}$\\ \hline \hline
4&46.1&25.9&2.77&4.85&3.84&2.08&0.932&0.183\\
5&103&32.5 (37)&4.04&5.69&4.88 (4.83)&1.65 (2.32)&0.943
(0.821)&0.181\\
6&190&39.1 (44)&5.21&6.58&5.91 (5.80)&1.37 (1.98)&0.949
(0.828)&0.179\\
7&313&45.7 (51)&6.33&7.50&6.92 (6.85)&1.17 (1.76)&0.952
(0.844)&0.177\\
8&477&52.2&7.42&8.44&7.93&1.02&0.954&0.174\\ \hline
\end{tabular*}
\end{center}
\caption
{Physical parameters within the vortex sheet
approximation for $\lambda = 1/2$ and $g=1000$. The values in
parentheses are obtained from the solution of the GP
equation~\cite{fetter05}. The last
column gives the vortex core radii as determined in Sec. IV. All
lengths are in units of $d_\perp$.}
\label{table}
\end{table}

\begin{figure}[h]
\scalebox{.6}
{\includegraphics[20,140][580,510]{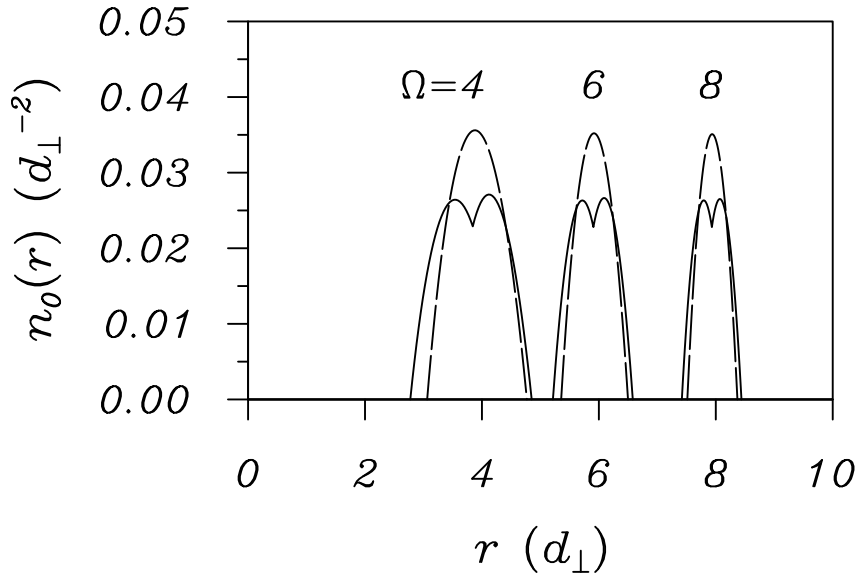}}
\caption{The solid curves show the density as a function of $r$ 
within the vortex sheet
approximation for three values of $\Omega$; $\lambda = 1/2$ and
$g=1000$. The dashed curves
are the densities for the giant vortex state. All quantities are
in units of $d_\perp$.}
\label{n_0}
\end{figure}
In Fig.~\ref{n_0} we show the equilibrium density $n_0(r)$ 
within the VSA for $g=1000$ for a few values of $\Omega$. 
The density exhibits a downward cusp at $r=R$ which corresponds to
the average depletion with respect to a smooth background  that the 
vortex cores give rise to. To make this more apparent, we have
also plotted the density for the giant vortex state which only
contains the central vortex. This shows the effect of
transferring circulation from the core of the giant vortex to
the vortex sheet near the middle of the annulus. 
The significant depletion of the density that occurs at
the vortex sheet radius is compensated by an increase in the 
width of the annulus. 
The fact that the depletion exhibits a cusp is of course an
artifact of treating the vortex sheet within a
Thomas-Fermi-like approximation. In reality,
the cusp will be smoothed out on a length scale $\xi$
characterizing the size of the vortex cores. Nevertheless, the
overall qualitative behaviour is expected to be a
good representation of the azimuthally averaged density in an
annulus containing a ring of vortices. We note further that there 
is a relatively slow recovery of the density from the region of
the sheet to the outer regions of the annulus. This is
associated with the long-range behaviour of the vortex sheet
velocity field. For a single vortex in a uniform gas of density
$n_{3D}$, the size of the core is set by the
healing length $\xi_0 = 1/\sqrt{8\pi a n_{3D}}$, but the 
density actually approaches its asymptotic value quite slowly:
$n(r) = n_{3D}(1-\xi_0^2/r^2 +\cdots)$. A similar
behaviour is occurring in the present context, but it is partially
masked by the finite width of the annulus. Another feature of
interest is the nearly constant maximum value of the density as
a function of $\Omega$. We can define an average density $\bar n$ in the
annulus by $2\pi Rd\bar n = 1$, where $d$ is the width
of the annulus. Since $R$ is approximately proportional to
$\Omega$ while $d$ is inversely proportional to $\Omega$, the
near constancy of $\bar n$, and hence the maximum, follows. 
The same argument also applies to the giant vortex state.

\begin{figure}[h]
\scalebox{.6}
{\includegraphics[20,140][580,510]{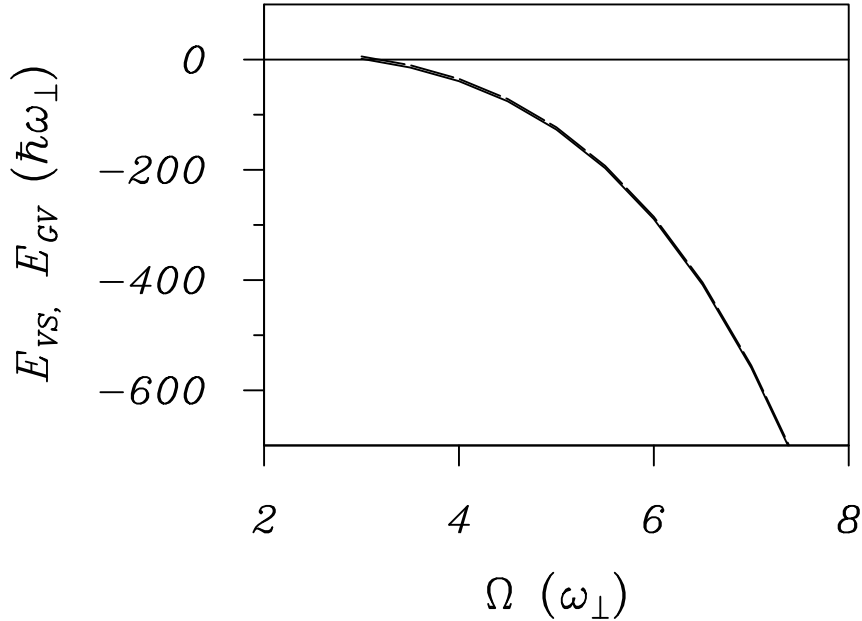}}
\caption{The solid curve gives the vortex sheet energy $E_{VS}$
and the dashed curve the giant vortex energy $E_{GV}$ (in units
of $\hbar \omega_\perp$), as a
function of the angular velocity $\Omega$ (in units of
$\omega_\perp$), for $g=1000$ and
$\lambda =1/2$.  }
\label{energy}
\end{figure}

The various parameters that emerge for the VS state are
collected in Table I. 
The asymptotic dependences $R \simeq \sqrt{(\Omega^2
-1)/2\lambda}$, $N_0\simeq \Omega^3/2\lambda$ and $N_r \simeq
5\Omega w_\infty/12$ are found to work quite well down to $\Omega =4$.
Also given in parentheses in the table are the values
of the parameters as determined by a numerical solution of the
Gross-Pitaevskii (GP) equation~\cite{fetter05}. 
We see that the VS approximation
underestimates the circulation of the GP solution by about 10\%
while the discrepancy in the radius is only a few percent. 
The discrepancy in the width $d$ of the annulus is much larger, with
the VS width approximately 30\% smaller than the GP result.
It should be noted, however, that there is some arbitrariness in the
way that the GP widths are extracted from the numerical data since the
GP densities fall off smoothly to zero~\cite{jackson}. 
In fact, despite its inherent uncertainty, visual inspection of the
published figures~\cite{fetter05} yields widths which
are in much closer agreement with our VS results.
We thus believe that the difference in the widths is actually
much smaller than indicated in Table I. Finally, the difference
between our inter-vortex spacings $b$ and the GP results
simply reflects the different values of the circulations $N_r$
in the two calculations. Interestingly, we see the same slight
increase of $b$ with increasing $\Omega$ as seen
previously~\cite{fetter05}.
We conclude that the VS approximation provides a reasonably 
faithful average representation of the state
containing a ring of vortices.
 
\begin{figure}[h]
\scalebox{.6}
{\includegraphics[20,140][580,510]{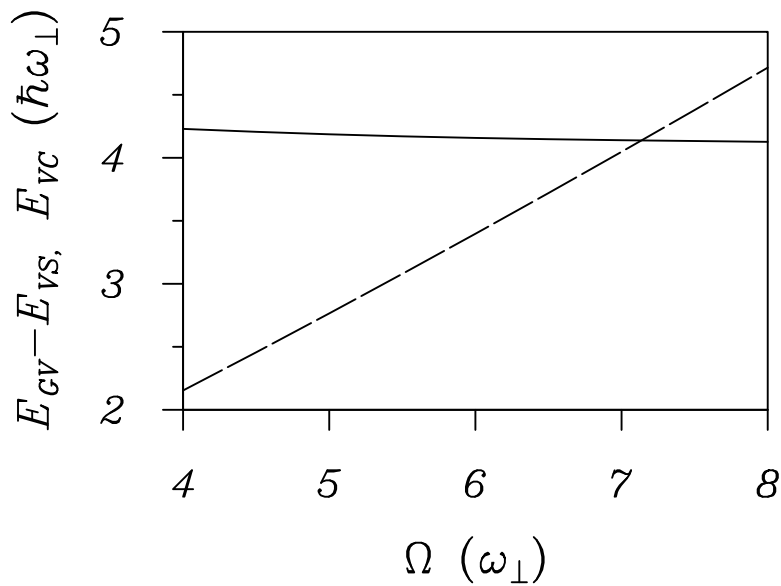}}
\caption{The solid curve is the difference, $E_{GV}-E_{VS}$,
between the giant vortex and vortex sheet energies, as a
function of the angular velocity $\Omega$, for $g=1000$ and
$\lambda =1/2$. The dashed curve is the vortex core energy
$E_{VC}$. The point of intersection of these two curves defines
the critical angular velocity $\Omega_c$ beyond which the giant
vortex state is more stable.}
\label{intersect}
\end{figure}
In Fig.~\ref{energy} we show $E_{VS}$ and $E_{GV}$ as a function of 
$\Omega$ for $g=1000$ and $\lambda=1/2$. To leading order in
$\Omega$, both of these energies behave as $-\Omega^4/8\lambda$.
The difference between them can be seen to be small and is
displayed in more detail in Fig.~\ref{intersect}.
We see that $E_{GV} - E_{VS} \simeq 4\hbar \omega_\perp$ over
the range of $\Omega$ shown, that is, the vortex sheet always
has a lower energy than the giant vortex. This is to be expected
since the insertion of a ring of vortices brings the velocity
field closer to that of rigid body rotation. Likewise, the
insertion of another vortex sheet would be expected to lower the energy
further. As we shall see, this is indeed the case. However, the
relative stability of these vortex sheet states is overstated
since the energy associated with the vortex cores has not been
included. The vortex core energy will at some point make states
with a large number of vortices less stable. In the
following section we calculate the vortex core energy to correct
the energy as determined in the VS approximation.
\section{Core Energy}
\label{core}
We begin by considering the first term in Eq.~(\ref{E_VC}) which
represents the quantum mechanical kinetic energy of the
condensate. The main contribution to this integral comes from
the region of the vortex cores which are defined by the envelope
function $F(\br)$. Using Eq.~(\ref{density}) and treating $\tilde n$ 
as a smooth function, the first term in Eq.~(\ref{E_VC}) gives
\begin{equation}
E_{VC}^{(1)} \simeq \int d^2r\, {\tilde n |\nabla F|^2 \over 8F} =
N_r \int_A  d^2r\, {\tilde n |\nabla F|^2 \over 8F}\,,
\end{equation}
where we have used the periodicity of $F$ to reduce the integral
to the area $A$ defined by $0 \le r < \infty$ and $-\pi/N_r \le
\phi \le \pi/N_r$. We assume that a vortex is positioned at
$r=R,\,\,\phi =0$ within this cell. If the core size $\xi$ is
small on the scale of the inter-vortex spacing, as turns out to be
the case, the
envelope function can be approximated within the cell as
\begin{equation}
F(\br) \simeq 1- e^{-\rho^2/\xi^2}\,,
\end{equation}
where $\rho$ is the distance from the vortex centre. Assuming
also that $\xi$ is small on the scale of the variations of
$\tilde n$, we find
\begin{equation}
E_{VC}^{(1)} \simeq C_1 N_r \tilde n(R)\,,
\label{EVC_1}
\end{equation}
where $C_1 = \pi(\pi^2/6 - 1)/2 = 1.01306\cdots$. It should be
noted that the dependence of this contribution on $\xi$ arises
solely through $\tilde n(R)$.

The next contribution to $E_{VC}$ is
\begin{eqnarray}
E_{VC}^{(2)} &=& {1\over 2} \int d^2r\, nv_1^2 \cr
&\simeq& {1\over 2} N_r \tilde n(R) \int_A d^2r
(1-e^{-\rho^2/\xi^2}) v_1^2 \,.
\label{EVC_2}
\end{eqnarray}
Using Eqs.~(\ref{v1r}) and (\ref{v1phi}), we have
\begin{eqnarray}
v_1^2(\br) 
&=& \left ( {N_r\over r} \right )^2 {\beta^2 \over 1-\beta^2}
\left [
1+ 2\sum_{k=1}^\infty \beta^k \cos kN_r\phi \right ] \cr
&=& \left ({N_r\over r}\right )^2 {\beta^2 \over 1+\beta^2 -2\beta 
\cos N_r\phi} \,.
\label{v1_squared}
\end{eqnarray}
The latter form is useful in order to see that $v_1^2 \propto
\rho^{-2}$ near each vortex. This singularity is cancelled by
the envelope factor in Eq.~(\ref{EVC_2}). However, in order to evaluate
$E_{VC}^{(2)}$ the first line in Eq.~(\ref{v1_squared}) proves to be 
more useful
despite the appearance of the factor $(1-\beta^2)^{-1}$ which is
singular at $r=R$. As we shall see, this singularity is
removable. 

To proceed we simplify the envelope factor using $\rho =
|\br-\bR| = \sqrt{r^2+R^2 - 2rR\cos\phi}\,$ (see Fig.~\ref{ring}):
\begin{eqnarray}
1-e^{-\rho^2/\xi^2} &=& 1-e^{-(r^2+R^2)/\xi^2}
e^{2rR\cos\phi/\xi^2}\cr
&\simeq& 1-e^{-(r-R)^2/\xi^2} e^{-rR\phi^2/\xi^2}\cr
&=& \big (1-e^{-(r-R)^2/\xi^2}\big ) + e^{-(r-R)^2/\xi^2}
\big (1-e^{-rR\phi^2/\xi^2} \big )\,. \nonumber
\end{eqnarray}
In going to the second line we have made use of the fact that
the angular range in $\phi$ is small when $N_r$ is large while
the grouping of terms in the last line is introduced to facilitate the
cancellation of singularities that arise. Accordingly, we
consider the integral
\begin{eqnarray}
I_1 &=& \int_A d^2r \big (1-e^{-(r-R)^2/\xi^2}\big ) v_1^2 \cr
&=& 2\pi N_r \int_0^\infty {dr\over r} \big
(1-e^{-(r-R)^2/\xi^2}\big )
{\beta^2\over 1-\beta^2} \,.
\label{I1_int}
\end{eqnarray}
It is apparent that the singularity at $r=R$ is cancelled by the
zero in the envelope factor. We show in Appendix A that this integral 
can be evaluated analytically to a good approximation, with the result
\begin{equation}
I_1 = 2\pi \left [ \ln\left ({aR\over N_r \xi}\right ) + 
{\sqrt{\pi}\over 2} N_r \left ({\xi \over R} \right ) 
- \bigg(M_2 + {1\over 2}N_r\bigg ) \left ( {\xi \over R} \right
)^2 + {\cal O}\left ({\xi\over R} \right )^3 \right ]\,,
\label{I_1}
\end{equation}
where $M_2 = {1\over 6}(N_r -{1\over 2})(N_r-{5\over 2})$ and $a
= {1\over 2}e^{\gamma/2} = 0.6673...$, with $\gamma$ being
Euler's constant~\cite{abramowitz70}.  The logarithm is associated with
the $(r-R)^{-1}$ singularity of the integrand in Eq.~(\ref{I1_int}) 
which is cut off at the distance $\xi$. The appearance of $N_r$ in the
logarithm indicates that $\xi/b$ is the physically relevant
parameter, with $b$ the inter-vortex spacing. This is also
true of the dominant power law terms in the expansion which is
given to higher order in Appendix A.

\begin{figure}[h]
\scalebox{.5}
{\includegraphics[40,120][590,585]{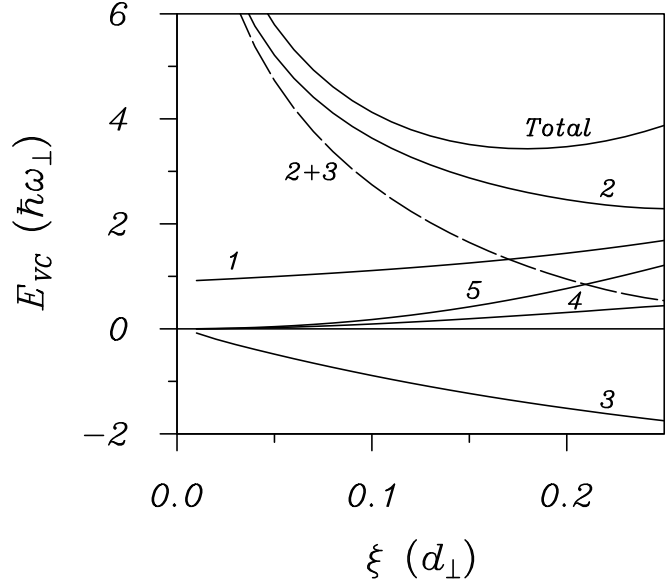}}
\caption{The total vortex core energy in units of
$\hbar\omega_\perp$ as a function of the core radius $\xi$ in
units of $d_\perp$, for $\Omega =6$, $g=1000$ and $\lambda =
1/2$. The numerically labelled 
curves indicate the various components $E_{VC}^{(i)}$, 
$i=1,\cdots,5$. The dashed curve gives the
combined result $E_{VC}^{(2)}+E_{VC}^{(3)}$, as this represents
the total change in classical kinetic energy with respect to the
VS state. The minimum in the total core energy gives the optimal
core radius.
}
\label{e_core}
\end{figure}

The second contribution to $E_{VC}^{(2)}$ involves the integral
\begin{eqnarray}
I_2 &=& \int_A d^2r\, e^{-(r-R)^2/\xi^2} (1-e^{-rR\phi^2/\xi^2}) v_1^2
\cr
&=&
N_r^2\int_0^\infty {dr\over r} e^{-(r-R)^2/\xi^2}
{\beta^2 \over 1-\beta^2} \int_{-\pi/N_r}^{\pi/N_r}
d\phi\,
\big (1-e^{-rR\phi^2/\xi^2}\big ) \left [
1+ 2\sum_{k=1}^\infty \beta^k \cos kN_r\phi \right ] \,.
\label{I2_int}
\end{eqnarray}
In this case all Fourier components of $v_1^2$ contribute. We
require the discrete Fourier transform of a gaussian which is
given by
\begin{equation}
2\pi g_m \equiv N_r \int_{-\pi/N_r}^{\pi/N_r} e^{-rR\phi^2/\xi^2} \cos
mN_r\phi\,d\phi = \sqrt{\pi \over
\kappa}e^{-m^2/4\kappa} \Re [{\rm erf}(z_m)]\,,
\label{g_m}
\end{equation}
where $\kappa = rR/\xi^2N_r^2$, $z_m = \pi\sqrt{\kappa} +
im/2\sqrt{\kappa}$ and ${\rm erf}(z)$ is the error
function~\cite{abramowitz70}.
The inverse Fourier transform is
\begin{equation}
e^{-\kappa \phi^2}  = \sum_{m = -\infty}^{\infty} g_m e^{im\phi}
\end{equation}
from which follows the useful identity
\begin{equation}
\sum_{m = -\infty}^{\infty} g_m = 1\,.
\label{identity}
\end{equation}
With these results the angular integral in Eq.~(\ref{I2_int}) 
can be expressed as
\begin{eqnarray}
N_r \int_{-\pi/N_r}^{\pi/N_r} (1-e^{-rR\phi^2/\xi^2} ) \left
[1+2\sum_{m=1}^\infty \beta^m \cos mN_r\phi \right ]\,d\phi&=&
2\pi - \sqrt{\pi \over\kappa} {\rm erf}(\pi\sqrt{\kappa}) -2
\sqrt{\pi \over\kappa}
\sum_{m=1}^\infty \beta^m e^{-m^2/4\kappa} \Re [{\rm erf}(z_m)]
\nonumber \cr &=& 2\sqrt{\pi \over\kappa}  
\sum_{m=1}^\infty (1-\beta^m) e^{-m^2/4\kappa} \Re [{\rm
erf}(z_m)]\,.\cr
\end{eqnarray}
Thus we find
\begin{equation}
I_2 = 2N_r \int_0^\infty {dr\over r} e^{-(r-R)^2/\xi^2} \beta^2
\sqrt{{\pi \over \kappa}}
\sum_{m=1}^\infty {1-\beta^m \over 1-\beta^2} e^{-m^2/4\kappa} 
\Re [{\rm erf}(z_m)]\,.
\label{I2}
\end{equation}
We now see that the $(1-\beta)^{-1}$ singularity is cancelled
by the numerator, leaving the factor
$\sum_{k=0}^{m-1}\beta^k/(1+\beta)$.
Both the gaussian and $\beta^2$ factors are highly localized
around $r=R$, which allows the sum to be evaluated at $r=R$ to a
good approximation. We thus find 
\begin{equation}
I_2 = C_2 N_r \int_0^\infty {dr\over r} e^{-(r-R)^2/\xi^2} \beta^2\,,
\label{I2_2}
\end{equation}
where
\begin{equation}
C_2 = \sqrt{{\pi \over \kappa}} 
\sum_{m=1}^\infty m  e^{-m^2/4\kappa} \Re [{\rm erf}(z_m)]\,.
\end{equation}
The integrals in Eq.~(\ref{I2_2}) on the ranges $(0 \le r \le R)$ and 
$(R \le r < \infty)$ are examples of the $J_k^{>/<}$ integrals 
defined in Appendix B. The constant $C_2$
can be expressed conveniently as the integral
\begin{equation}
C_2 = 2\kappa \int_0^{\pi} \phi \cot{\phi \over 2}
e^{-\kappa\phi^2}\, d\phi\,,
\end{equation}
which shows that $C_2$ behaves asymptotically for large
$\kappa$ as $C_2 \simeq 2\sqrt{\pi\kappa}(1-{1\over 24 \kappa}+
\cdots)= 2\sqrt{\pi}R/\xi N_r + \cdots$. Since the integral in
Eq.~(\ref{I2_2}) is proportional to $\xi$ as $\xi \to 0$, $I_2$
itself has a finite limiting value in this limit, unlike the
logarithmically divergent behaviour of $I_1$. 
Our final result for $E_{VC}^{(2)}$ is
\begin{equation}
E_{VC}^{(2)} = {1\over 2} N_r \tilde n(R)(I_1+I_2)\,.
\end{equation}
This quantity is plotted as a function of $\xi$ in Fig.~\ref{e_core}.

The next contribution to the core energy is
\begin{eqnarray}
E_{VC}^{(3)} &=& \int d^2r \,nv_0v_{1\phi} \cr
&\simeq& -N_r\tilde n(R) \int_A d^2r \, e^{-(r-R)^2/\xi^2}
e^{-rR\phi^2/\xi^2} v_0 v_{1\phi}\,.
\label{EVC_3}
\end{eqnarray}
This is the cross term that arises from squaring $\bv_0+\bv_1$.
For $r>R$,
$v_0$ and $v_{1\phi}$ have the same sign, indicating that the
velocity is actually larger than $v_0$ close to the sheet. On
the other hand, $v_{1\phi}$ has the opposite sign for $r<R$,
implying that there is a cancellation between $v_0$ and $v_{1\phi}$ on 
this side of the sheet.
The overall negative sign in Eq.~(\ref{EVC_3}) is due to the
fact that only the regions within the vortex cores contribute to
the integral. In
these regions the density is depleted with respect to the average
$n_0$ and $E_{VC}^{(3)}$ compensates for the contribution these regions
make to the VSA kinetic energy.

The angular integral in Eq.~(\ref{EVC_3}) is the same as 
encountered previously in
the calculation of $E_{VC}^{(2)}$. We have
\begin{equation}
E_{VC}^{(3)} = -N_r\tilde n(R) \int_0^\infty dr\,
e^{-(r-R)^2/\xi^2} v_0(r) {\rm sgn}(r-R)\sqrt{{\pi \over \kappa}}
\sum_{m=1}^\infty \beta^m e^{-m^2/4\kappa} \Re[{\rm erf}(z_m)]\,.
\end{equation}
We retain only the rapidly varying $\beta^m$ factor from the sum
in doing the radial
integral and set $r=R$ in the remaining slowly varying factors.
We then obtain
\begin{equation}
E_{VC}^{(3)} = -N_r\tilde n(R) \sqrt{{\pi \over \kappa}}
\sum_{m=1}^\infty \left [ (N_0+N_r) J_{mN_r+1}^>(R/\xi) - N_0
J_{mN_r-1}^<(R/\xi) \right ] e^{-m^2/4\kappa} \Re[{\rm erf}(z_m)]\,,
\end{equation}
where the $J_k^{>/<}$ integrals are defined in Appendix B. Within
the sum there is a strong cancellation between the two $N_0$ 
terms, leaving mainly the term proportional to $N_r$. This is most
evident in the $\xi \to 0$ limit for which the $J_k^{>/<}$ integrals
take a limiting value of $\sqrt{\pi}\xi/2R$. The remaining sum
is evaluated using Eq.~(\ref{identity}), giving 
$E_{VC}^{(3)} \simeq -N_r \tilde n(R)\pi^{3/2}(N_r\xi/2R)$
to lowest order in $\xi$. This shows
that $E_{VC}^{(3)}$ depends linearly on $\xi$ for $\xi \to 0$.

The contribution
$E_{VC}^{(4)}=-\int d^2r \Omega r n v_{1\phi}$ is essentially of
the same form as $E_{VC}^{(3)}$ and can be handled in a 
similar way. We find
\begin{equation}
E_{VC}^{(4)} = N_r\Omega R^2 \tilde n(R) \sqrt{{\pi \over \kappa}}
\sum_{m=1}^\infty \left [ J_{mN_r-1}^>(R/\xi) -
J_{mN_r+1}^<(R/\xi) \right ] e^{-m^2/4\kappa} \Re[{\rm erf}(z_m)]\,.
\end{equation}
We recall that $\Omega R^2 = N_0 + N_r/2$, with $N_0$ typically
being much larger than $N_r$. However, the cancellation between
$J_{mN_r-1}^>$ and $J_{mN_r+1}^<$ within the sum diminishes the
size of this contribution in comparison to $E_{VC}^{(3)}$, as can
be seen in Fig.~\ref{e_core}.

Finally, we have the interaction term
\begin{eqnarray}
E_I &=& {1\over 2} g \int d^2r\, n^2 \cr
&=& {1\over 2} N_r g \int_A d^2r [1-h(\br)]^2 \tilde
n^2(r)\cr
&=& {1\over 2} g \int d^2r n_0^2 + {1\over 2} g \int
d^2r \big [ (h^2)_0- (h_0)^2 \big ] \tilde n^2(r)\,,
\label{E_I}
\end{eqnarray}
where $h(\br) = \exp{(-(r-R)^2/\xi^2)}\exp{(-rR\phi^2/\xi^2)}$
and the subscript 0 in the second term
indicates the zeroth Fourier component as
defined in Eq.~(\ref{g_m}). The first term in Eq.~(\ref{E_I})
contributes to $E_{VS}$, while the second term represents
the interaction contribution to $E_{VC}$. We thus find
\begin{eqnarray}
E_{VC}^{(5)} &=& {1\over 2} g \int d^2r 
\big [(h^2)_0 - (h_0)^2 \big ]
\tilde n^2(r) \nonumber \cr
&=& {1\over 2} g \int_0^\infty dr \,r e^{-2(r-R)^2/\xi^2}
\left [ \sqrt{{\pi\over 2\kappa}}{\rm erf}(\pi\sqrt{2\kappa})
 - {1 \over 2\kappa} {\rm erf}^2(\pi\sqrt{\kappa}) \right ]
\tilde n^2(r)   \nonumber \cr
&\simeq& {\pi\over 4} g N_r \xi^2 \tilde n^2(R) \left [ {\rm
erf}(\pi\sqrt{2\kappa}) - {1\over \sqrt{2\pi\kappa}} {\rm
erf}^2(\pi\sqrt{\kappa}) \right ]\,,
\end{eqnarray}
with $\kappa = (R/N_r\xi)^2$.\hfil

In Fig.~\ref{e_core} we show the various contributions to the 
core energy as a function of $\xi$, together with the total 
$E_{VC} = \sum_{i=1}^5 E_{VC}^{(i)}$, for the case $g=1000$ 
and $\Omega=6$. The
behaviour for other values of these parameters is very similar.
All the contributions except for
$E_{VC}^{(3)}$ are seen to be positive. As explained previously,
the latter is the correction to the kinetic energy that arises
from the interference between $v_0$ and $v_{1\phi}$. Since
$E_{VC}^{(2)}$ and $E_{VC}^{(3)}$ both represent corrections to
the VS kinetic energy, we have also plotted the sum of these two
terms. The combination is seen to decrease monotonically with
increasing $\xi$. The other contributions $E_{VC}^{(1)}$,
$E_{VC}^{(4)}$ and $E_{VC}^{(5)}$ are positive and increase with
$\xi$. The total core energy $E_{VC}$ exhibits a minimum 
at $\xi_{\rm min} \simeq 0.179$ which is the equilibrium
core radius for the case being considered. Values of $\xi_{\rm
min}$ for other $\Omega$ are given in Table I. The core radius is
seen to be rather insensitive to the rotation rate. These values are
of the order of, but somewhat larger than, the local bulk healing 
length $\xi_0 \simeq 1/\sqrt{8\pi a \sigma \tilde n(R)} \simeq 0.12$.

\begin{figure}[h]
\scalebox{.6}
{\includegraphics[20,140][580,510]{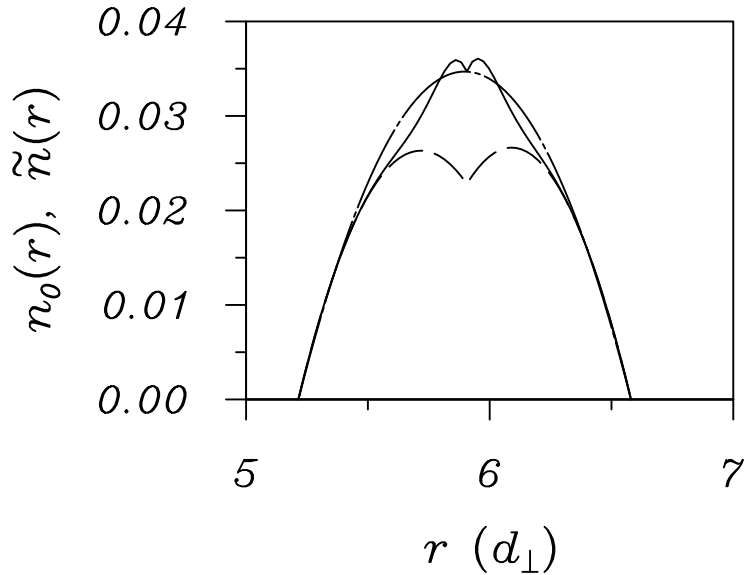}}
\caption{The solid curve shows the background density, $\tilde
n(r)=n_0(r)/F_0(r)$, as a function of $r$ for $\Omega =6$, $g=1000$ and
$\lambda =1/2$. The dashed curve is the density in the vortex
sheet approximation, $n_0(r)$. The chain curve is a possible
refined smooth background as discussed in the text.}
\label{ntilde_fig}
\end{figure}

The dependence on $\xi$ of the quantum kinetic energy term
$E_{VC}^{(1)}$ is entirely due to the $\tilde n(R)
=n_0(R)/F_0(R)$ factor, which also appears in all the other core
energy contributions. The angular average of
the envelope function is given by $F_0(R) = 1-{\rm
erf}(\pi\sqrt{\kappa})/2\sqrt{\pi\kappa}$ and its dependence on
$\xi$ appears through $\kappa =
(R/N_r \xi)^2$. For $\kappa > 1$ the error function is
approximately unity and $F_0(R)^{-1} \simeq 1 +
\sqrt{\pi}(\xi/b) + \pi (\xi/b)^2+\cdots$ where $b=2\pi R/N_r$ is
the inter-vortex spacing. At the minimum, $\xi_{\rm min}/b\simeq 0.2$ 
and the $F_0(R)^{-1}$ factor makes $\tilde n(R)$
about 50\% larger than $n_0(R)$. This is perhaps a slight overestimate
of the background density since the gaussian envelope
function gives a core density profile that is more compact than what
one would expect it to actually be.
As explained earlier, the core density
recovers its asymptotic value rather slowly due to the slow decay
of the azimuthally averaged velocity. This latter effect is accounted
for in an average way within the VSA but is not accounted for using the
gaussian core profile. In Fig.~\ref{ntilde_fig} we compare 
$n_0(r)$ and $\tilde n(r)$; the enhancement of $\tilde n(r)$
above $n_0(r)$ is evident. However the detailed structure in
$\tilde n(r)$ should not be taken seriously. The cusp at $r = R$
is of course a residual artifact of the cusp in
$n_0(r)$; it would be eliminated if the smooth core profile were
accounted for in $n_0(r)$. Apart from this, the
figure suggests that $\tilde n(r)$ would appear
smoother if the gaussian where augmented by wings
which decayed more slowly so as to be more consistent with the
long-range part of the core density profile contained in
$n_0(r)$. To give an impression of what an improved $\tilde n(r)$ might
look like, we have plotted the function $\tilde
n_{\rm fit}(r) = A(r-R_1)(R_2-r)$ with $A$ chosen to reproduce
$\tilde n(R)$. With this choice of $A$ one can see that $\tilde
n_{\rm fit}(r)$ goes smoothly to $n_0(r)$ at the edges of the
annulus and appears to be a reasonable candidate for the true smooth 
background. The ratio $n_0(r)/\tilde n_{\rm fit}(r)$ can be
thought of as a refined $F_0(r)$ which in turn implies a refined
core density profile. It should be emphasized, however, that
the detailed shape of $\tilde n(r)$ is not particularly relevant
since only $\tilde n(R)$ is used in estimating the vortex core
energy. In view of the variational nature of the calculation,
refinements in the core density profile would not be expected to
lead to significant changes in the value of the core energy.

\section{Results}

We now combine the core energy with the VS energy calculated in
the Sec. III. To begin, we do this perturbatively, that
is, we assume that the core energy is a small correction to the
VS energy. This is certainly correct for large $\Omega$ but will become
less accurate as $\Omega$ is reduced. Nevertheless, for the time being
we use the parameters $R$, $N_r$ and $n_0(R)$ which arise from the 
minimization of $E_{VS}$ to evaluate $E_{VC}$ over the whole
range of $\Omega$. The procedure provides an
upper bound to the total energy $E_{VS}$$+$$E_{VC}$ of the annular array
and therefore underestimates its stability relative to the giant 
vortex state. To determine the transition point we plot
$E_{GV}$$-$$E_{VS}$ and $E_{VC}$ versus $\Omega$. 
This is done in Fig.~\ref{intersect} for the case of
$g=1000$. As stated earlier, $E_{GV}$$-$$E_{VS}$ is almost constant as a
function of $\Omega$, while $E_{VC}$ increases approximately linearly.
This latter behaviour is due to the fact that all the vortex core
energies are proportional to $N_r$ which itself is approximately a 
linear function of $\Omega$. It should be noted that the
vortex core contributions also depend on the parameter $N_r\xi/R$, but
this parameter is approximately constant since $R$ is also proportional
to $\Omega$. 

\begin{figure}[h]
\scalebox{.6}
{\includegraphics[20,140][580,510]{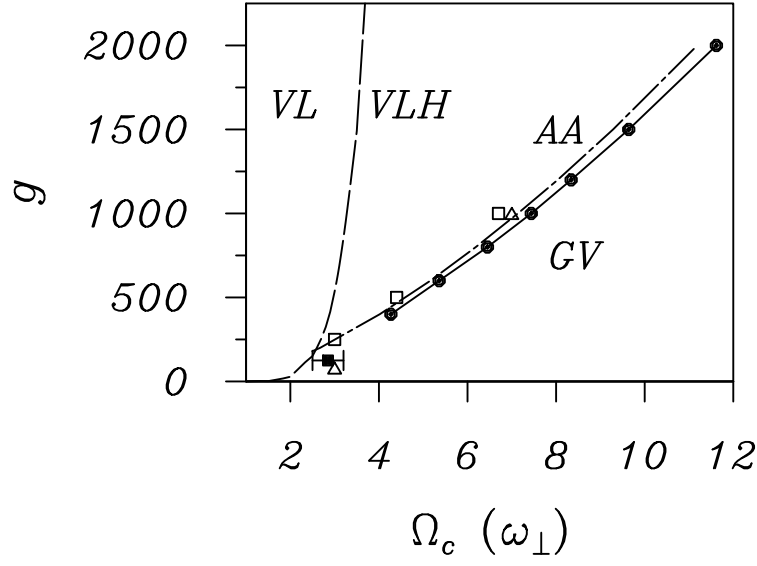}}
\caption{The dash-dot curve shows the phase boundary between the
annular array (AA) and the giant vortex (GV) state as determined
by treating the core energy perturbatively. The solid points
joined by solid lines
denote the position of the phase boundary as determined by a
global minimization of the total energy. The open triangles are
the values determined by the GP solution~\cite{fetter05} (the
$g=1000$ point is a lower bound for $\Omega_c$) 
and the open squares are the
results of Kim and Fetter~\cite{kim05}. The filled square with
error bars gives the approximate bounds on $\Omega_c$
as determined in Ref.~\cite{kasamatsu02} for $g=125$.
The dashed curve denotes the
onset of a hole (VLH) in the vortex lattice (VL).
}
\label{phase}
\end{figure}
The intersection in Fig.~\ref{intersect} occurs at the angular
velocity $\Omega_c \simeq 7.1$. 
Repeating the calculations as a function of $g$ yields the phase
boundary between the vortex array and giant vortex states. These
results are presented as the dash-dot curve in Fig.~\ref{phase}
and are very similar to those obtained by Kim and Fetter~\cite{kim05} 
as indicated by the open squares.  These 
authors do not invoke the VS approximation but rather represent the
density as in Eq.~(\ref{density}), choosing the envelope function to be
composed of linear cores and the background density $\tilde n(r)$ to
have an analytic form that was found to work well for the giant vortex. 
They then fix the number of vortices in
the ring, $N_r$, and minimized the GP energy with respect to the
remaining parameters. They find a local minimum in this energy as a 
function of $N_r$ which corresponds to the annular array and then 
compare this energy with that of the giant vortex ($N_r = 0$).
Remarkably, the minimum values of $N_r$ determined by this approach
agree very well with our VS results in Table~\ref{table}, 
even though the latter excludes
our core correction. However it should be emphasized that the VS
approximation already includes the vortex cores in an average sense,
which is why our VC energy is typically a relatively small correction.
For $g= 1000$ and $\lambda = 1/2$ they find $\Omega_c \simeq 6.7$,
slightly smaller than our perturbative result of 7.1. The fact that our
critical angular velocity is slightly higher would suggest
that our variational
treatment of the annular array is somewhat better, but given the
differences in the approaches, the two results should be considered as 
being essentially in agreement. Their results in fact approach
ours for smaller values of $g$.

\begin{figure}[h]
\scalebox{.6}
{\includegraphics[54,185][565,535]{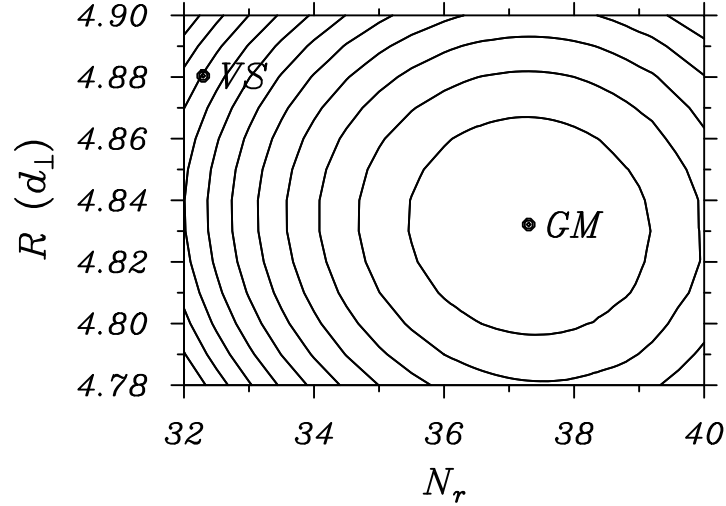}}
\caption{Total energy contours ($E_{VS} + E_{VC}$) in the
vicinity of the global minimum (GM). The point labelled VS
is the location of the minimum in the vortex sheet approximation 
($E_{VS}$). The parameters in the calculation are $\Omega =5$,
$g=1000$ and $\lambda = 1/2$.
}
\label{contour}
\end{figure}
A more accurate calculation would involve minimizing the total vortex
array energy ($E_{VS}+E_{VC}$) 
with respect to the two parameters $\nu_1$ and $\nu_2$, or
alternatively, $N_r=\nu_2-\nu_1$ and $R=\sqrt{(\nu_1+\nu_2)/2\Omega}$.
In Fig.~\ref{contour} we present a contour plot showing the
behaviour of the total energy in the vicinity of the global
minimum. It can be seen that $N_r$ and $R$ are almost orthogonal
variables. The point labelled $VS$ is the position of the minimum of
the vortex sheet energy $E_{VS}$ itself; constant energy
contours of $E_{VS}$ would appear similar to those shown but
would be centred on $VS$. As one moves away from $VS$ along a line at
constant $R$, $E_{VS}$ of course increases but the total
energy actually decreases as a result of the decrease in $E_{VC}$.
The fact that $E_{VC}$ decreases with $N_r$ is somewhat surprising 
since $E_{VC}$ is explicitly
proportional to $N_r$. However, $E_{VC}$ also depends on the parameter
$\xi/b$, and the inter-vortex spacing is decreasing with increasing
$N_r$. This dependence is evidently dominating the variation of
$E_{VC}$ with $N_r$.

To locate the global minimum we have found it convenient to
begin by using the fixed-$N_r$ minimization scheme in
Eq.~(\ref{N_r-fixed}) which determines the minimum value of 
$E_{VS}$ for a given value of $N_r$. $R$ is found to
vary only slightly as $N_r$ is varied and the minimum of the
total energy along this line in the $N_r$-$R$ plane can readily
be determined. With $N_r$ fixed at the value where this minimum
occurs, we subsequently evaluate the total energy as a function of $R$,
adjusting the chemical potential to ensure that Eq.~(\ref{g_norm}) 
is satisfied. The global minimum can then be approached quickly
with successive variations of $N_r$ and $R$. In this way we find a
global minimum at $N_r = 37.5$ and $R=4.83$ for $\Omega=5$.
These values are perhaps coincidentally close to the GP results 
of 37 and 4.83, respectively; the agreement with the GP values
(in brackets) is slightly poorer for $\Omega =6$ where we find
$N_r = 47.0$ (44) and $R=5.87$ (5.80) and for
$\Omega =7$, $N_r = 57.5$ (51) and $R=6.88$ (6.85).
Nevertheless, the global minimization in all cases moves $N_r$ in 
the right direction from the VS values in Table I, and yields 
values of $R$ that are in excellent agreement with the GP
results. It is therefore clear that our variational approach is
providing a good description of the annular array properties at
this value of $g$.

By repeating these calculations as a function of $\Omega$ and
comparing the energy at the global minimum with $E_{GV}$, we
obtain an improved value of $\Omega_c$. These points are plotted 
in Fig.~\ref{phase} and show that the phase boundary shifts
to slightly larger angular velocities; the fractional change in
$\Omega_c$ is 3-6\% and increases with decreasing $g$. The phase
boundary for the global minimum ends at $g=200$; below this
interaction strength the
total energy did not exhibit a local minimum
corresponding to the annular array. 

\begin{figure}[h]
\scalebox{.6}
{\includegraphics[20,140][580,510]{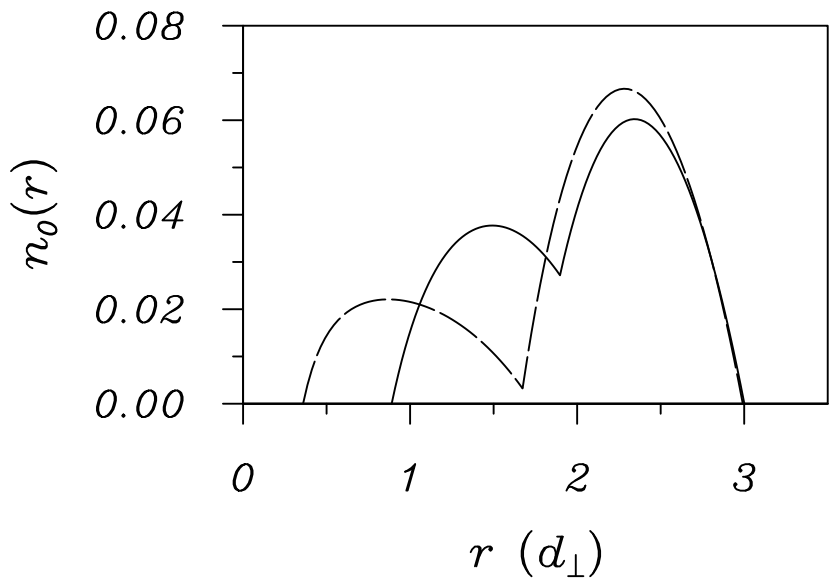}}
\caption{The density in the vortex sheet approximation for ($N_r
= 10$, $N_0 = 4$), solid curve, and ($N_r= 12$, $N_0 = 1$),
dashed curve. The results are obtained for
$\Omega = 2.5$, $\lambda = 1/2$ and $g=125$.
}
\label{kasamatsu}
\end{figure}
We believe this apparent failure of the calculations at low
values of $g$ is associated with the breakdown of the
Thomas-Fermi approximation used for the vortex sheet. To confirm
this we have analyzed in more detail the case of
$g=125$ at $\Omega=2.5$ which was studied previously by
Kasamatsu {\it et al.}~\cite{kasamatsu02}. Here, we calculated the
total energy as a function of integral values of $N_0$ and
$N_r$. With $N_r$ fixed at 10, we find a minimum energy at $N_0=4$ which
matches the ($N_r =10$, $N_0=4$) equilibrium 
state found in Ref.~\cite{kasamatsu02}.
The VS density for this state shown in Fig.~\ref{kasamatsu}
corresponds quite well to the density given in Fig.~1 of
Ref.~\cite{kasamatsu02} with regard to the boundaries of the
annulus, the radius of the vortex array and the relative density
of the inner and outer portions of the annulus. However, the
total energy we find is higher than that of the giant vortex
state ($-1.35\, \hbar \omega_\perp$ vs. $-1.41 \,\hbar
\omega_\perp$); according to our calculations, 
$\Omega = 2.5$ lies to the right of the phase boundary,
consistent with the extrapolation of the dot-dash curve in
Fig.~\ref{phase}, but not consistent with the stability of the
(10, 4) state as determined by the GP equation. As stated
earlier, we attribute this
discrepancy in $\Omega_c$ to the breakdown of the TF
approximation for the VS. In fact we
find states with even lower total energy when the central circulation 
is minimized. For $N_r =
12$ we find a minimum of $E=-1.5\, \hbar \omega_\perp$ at $N_0= 1$
and the corresponding density is shown by the dashed curve
in Fig.~\ref{kasamatsu}. It is clear from this figure why the 
energy of this state is lower in our approach. 
Although the VS energy is higher than for the (10, 4)
state, the vortex core energy $E_{VC}$ of the (12, 1) state 
is reduced to an even greater extent because of the 
low density $n_0(R)$ at the radius of the vortex sheet. (Recall
that the VC energy is proportional to this quantity.) However
we expect that there will be significant beyond-TF corrections
to the VS energy for this state due to the highly inhomogeneous
nature of the density. We believe this is the reason why the
(10, 4) state, as opposed to the (12, 1) state, is the lowest energy 
state as found in the GP calculations~\cite{kasamatsu02}.
We conclude that beyond-TF corrections must be playing a role in our
determination of the phase boundary for small values of $g$.

\begin{figure}[h]
\scalebox{.6}
{\includegraphics[20,140][580,510]{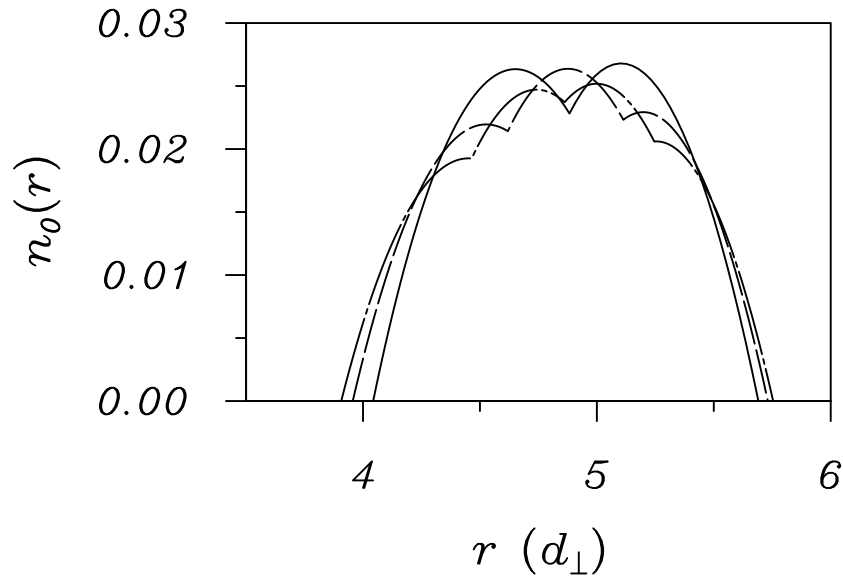}}
\caption{The density in the vortex sheet approximation for one
(solid), two (dash) and three (dash-dot) vortex sheets. The
results are obtained for $\Omega = 5$, $\lambda = 1/2$ and
$g=1000$.
}
\label{rings}
\end{figure}
We finally present some results for configurations involving 
multiple rings of vortices. These results are
obtained using a straightforward generalization of the vortex
sheet approximation discussed in Sec.~\ref{sheet} for a single ring.
In Fig.~\ref{rings} we show the density $n_0(r)$ obtained by
minimizing the VS energy functional 
for one, two and three vortex sheets for the case of $\Omega = 5$.
The energies for these cases are $-127.0$, $-128.2$ and $-128.7$
$\hbar\omega_\perp$, respectively,
indicating that the vortex sheet approximation prefers multiple
ring configurations. In fact, one can consider the limit of a
continuous distribution of vortex sheets with a circulation
density $\nu(r)$ which is a continuous function of $r$. The
distribution which minimizes the free energy is $\nu(r) =
\Omega /\pi$, that is, a constant density throughout the annular
region of the condensate. This gives rise to the rigid body
velocity field $v_{\rm RB}(r) = r\Omega$ as used in the
uniform vortex lattice calculations~\cite{fetter05}. 
Of course the vortex core energies must
be taken into account in order to determine the relative
stability of the multiple vortex sheet states. When this energy
is included we expect to find a
sequence of phase boundaries between the $n$ and $(n+1)$ sheet
states, and in particular, a phase boundary between the single
and double ring configurations which lies to the left of the
phase boundary shown in Fig.~\ref{phase}. 
Our calculation of the vortex core energy can be
extended to treat these multiple ring configurations but we will
not pursue this extension here.
\section{Conclusions}

In summary, we have investigated the properties of an annular
BEC in an harmonic plus quartic trap in the regime of high angular
velocities $\Omega > \omega_\perp$. Of particular interest is the 
transition from the state containing an annular array of vortices 
to the giant vortex state in which the circulation is carried by a
single central vortex. The phase boundary defining the transition 
between these two states was determined by a variational
analysis of the Gross-Pitaevskii energy functional.
Our approach identified two
contributions to the total energy. One, based on an azimuthal
average of the density and velocity, defined what we refer to as
the vortex sheet approximation. This part accounted for most of
the energy but important corrections to the energy coming from
the vortex cores had to be included in order to properly describe
the transition to the giant vortex state. The vortex core energy
was also treated variationally by assuming the core to have a
gaussian density profile. With this choice of the core profile we
were able to provide analytic expressions for
the various contributions to this part of the energy. The phase
boundary determined by treating the core energy perturbatively
differed only slightly from the phase boundary determined by a
global minimization of the total energy. However, the global
minimization did provide equilibrium parameters which were in
better agreement with those obtained from the solution of the
GP equation~\cite{fetter05}. Our results for the phase
boundary are expected to be less reliable in the limit of weak
interactions ($g < 250$) where corrections to our
Thomas-Fermi-like approximation to the vortex sheet appear to be
important. Our approach can also be used to deal with multiple
ring configurations and possibly with other arrangements of vortices
in rotating trapped gases.

As a final comment, we emphasize that we have only addressed
the static equilibrium properties of the vortex structures
considered. The dynamic stability of these states is also of
interest and was addressed in the paper by Kim and
Fetter~\cite{kim05}. Their analysis leads to the conclusion that
the vortices in the annular array are indeed dynamically stable.
Thus the annular array is a well defined equilibrium state that
is separated by a meaningful phase boundary from the giant
vortex state. 
\appendix
\section{}

We present here details of the calculation leading to the result
given in Eq.~(\ref{I_1}). The contribution to the integral in
Eq.~(\ref{I1_int}) from the range $R\le r < \infty$ can be written as
\begin{equation}
I_1^> = 2\pi N_r \int_1^\infty {dx \over x} \left
(1-e^{-(R/\xi)^2(x-1)^2} \right ) {1\over x^{2N_r}-1}
\end{equation}
Integrating by parts we have
\begin{equation}
I_1^> = {2\pi R^2 \over \xi^2} \int_1^\infty dx (x-1)
e^{-(R/\xi)^2(x-1)^2}  \ln \left ({x^{2N_r} \over x^{2N_r}-1}
\right )
\end{equation}
The $x^{2N_r}$ factor in the logarithm gives the integral (with $x =
1+y$)
\begin{eqnarray}
A &=& {4\pi N_r R^2 \over \xi^2} \int_0^\infty dy \, 
e^{-R^2y^2/\xi^2} y\ln(1+y) \nonumber \cr
&\simeq& {2\pi N_r \xi \over R} \left [ {\sqrt{\pi}\over 2}
-{1\over 2} {\xi \over R} +  {\sqrt{\pi}\over 4} \left ( {\xi \over
R}\right )^2 -{1\over 2} \left ( {\xi\over R} \right )^3
+\cdots \right ]\cr
\end{eqnarray}
The other logarithmic term is expanded as 
\begin{equation}
\ln(x^{2N_r} -1) \simeq \ln (2N_r y) + M_1 y + M_2 y^2 + \cdots
\end{equation}
and gives the contribution
\begin{equation}
B = - {2\pi R^2 \over \xi^2} \int_0^\infty dy \,
e^{-R^2y^2/\xi^2}  y \left [ \ln(2N_ry) + M_1 y + M_2 y^2 +
\cdots \right ]
\end{equation}
Similarly, the range $0\le r \le R$ gives the integral
\begin{eqnarray}
I_1^< &=& 2\pi N_r \int_0^1 {dx \over x} \left
(1-e^{-(R/\xi)^2(1-x)^2} \right ) {x^{2N_r} \over 1- x^{2N_r}}
\nonumber \cr
&=& - {2\pi R^2 \over \xi^2} \int_0^1 dx \,(1-x)
e^{-(R/\xi)^2(1-x)^2}  \ln \left (1- x^{2N_r} \right )
\nonumber \cr
&\simeq& - {2\pi R^2 \over \xi^2} \int_0^\infty dy \,
e^{-R^2y^2/\xi^2}  y \left [ \ln(2N_ry) - M_1 y + M_2 y^2 +
\cdots \right ]
\end{eqnarray}
where the assumption that $R/\xi$ is large allows the upper
limit to be extended to infinity. Adding $I_1^<$ to $I_1^> = A+B$ 
we finally obtain
\begin{equation}
I_1 = 2\pi \left [ \ln \left ( {aR\over N_r \xi} \right ) +
N_r {\sqrt{\pi}\over 2} {\xi \over R} - \left (M_2 + {N_r\over
2}\right ) \left (
{\xi \over R} \right )^2 + N_r {\sqrt{\pi} \over 4} \left ({\xi
\over R} \right )^3 - \left (2 M_4 + {N_r\over 2}\right ) \left 
( {\xi \over R} \right)^4 + \cdots\right ]
\end{equation}
where $a={1\over 2} e^{\gamma/2}$ with $\gamma$ equal to Euler's
constant, $M_2 = {1\over 6}(N_r-{1\over 2})(N_r-{5\over 2})$ and
$M_4 = -{1\over 1440}(N_r-{1\over 2})(8N_r^3+4N_r^2-218N_r+251)$. 
This expansion provides an accurate representation of
$I_1$ over the range of $\xi/R$ of interest.

\section{}
In Sec.~\ref{core}, we had integrals of the form
\begin{equation}
\int_R^\infty {dr \over r} e^{-(r-R)^2/\xi^2} (R/r)^{(k-1)} =
\int_0^\infty e^{-R^2y^2/\xi^2} (1+y)^{-k}\,dy \equiv
J_k^>(R/\xi)\,.
\end{equation}
For large $R/\xi$ we have to
a good approximation
\begin{eqnarray}
J_k^>(R/\xi) &=&  \int_0^\infty
e^{-R^2y^2/\xi^2} e^{-k\ln(1+y)}\,dy\nonumber \cr
&\simeq&  \int_0^\infty e^{-R^2y^2/\xi^2}
e^{-k(y-y^2/2)}\,dy \nonumber \cr
&=& {1\over 2} \sqrt{{\pi \over \alpha_-}}\,e^{\alpha_- y_-^2}
{\rm erfc}(\sqrt{\alpha_-} y_-) \cr
\end{eqnarray}
where $\alpha_- = (R/\xi)^2 - k/2$ and $y_- = k/2\alpha_-$.

The other integral that appears is
\begin{equation}
\int_0^R {dr \over r} e^{-(r-R)^2/\xi^2} (r/R)^{(k+1)} =
\int_0^1 e^{-R^2y^2/\xi^2} (1-y)^k\,dy\equiv J_k^<(R/\xi)\,.
\end{equation}
For large $R/\xi$ this becomes
\begin{eqnarray}
J_k^<(R/\xi) &=& \int_0^1
e^{-R^2y^2/\xi^2} e^{-k\ln(1-y)}\,dy\nonumber \cr
&\simeq& \int_0^\infty e^{-R^2y^2/\xi^2}
e^{-k(y+y^2/2)}\,dy \nonumber \cr
&=& {1\over 2} \sqrt{{\pi \over \alpha_+}}\,e^{\alpha_+ y_+^2}
{\rm erfc}(\sqrt{\alpha_+} y_+) \cr
\end{eqnarray}
where $\alpha_+ = (R/\xi)^2 + k/2$ and $y_+= k/2\alpha_+$.
\begin{acknowledgments}
One of us (E.Z.) would like to acknowledge useful discussions
with A.L. (Sandy) Fetter and Natasha Berloff. The hospitality of
the Aspen Center for Physics where part of this work was
performed is also gratefully acknowledged. This work was
supported by a grant from NSERC of Canada.
\end{acknowledgments}


\begin{references}
\bibitem{footnote1} This procedure is equivalent to that adopted, for
example, in Refs.~\cite{fischer03} and~\cite{kavoulakis03}.
\end{references}
\end{document}